# Cargo Delivery to Cells Using Laser-Irradiated Carbon-Black-Loaded PDMS

Weilu Shen*[1], Anqi Chen*[1], Gurminder K. Paink[1], Nicole Black[1], David Weitz[1,2], Eric Mazur[1,2]

[1]*John A. Paulson School of Engineering and Applied Sciences, Harvard University, Cambridge, MA 02138, USA*

[2]*Department of Physics, Harvard University, 9 Oxford Street, Cambridge, Massachusetts 02138*

*Authors contributed equally

**Abstract:** Effective intracellular delivery is essential for successful gene editing of cells. Spatially selective delivery to cells that is simultaneously precise, consistent, and non-destructive remains challenging using conventional state-of-the-art techniques. Here, we introduce a carrier-free method for spatiotemporal delivery of fluorescently labeled cargo into both adherent and suspension cells using carbon-black-embedded polydimethylsiloxane (PDMS) substrates irradiated by nanosecond laser pulses. This low-cost, biocompatible material, coupled with an optical approach, enables scalable, spatially selective, and sequential delivery of multiple cargo molecules, including FITC-dextran and siRNA, to a broad range of cells. Notably, we achieved siRNA delivery into the cytoplasm of hard-to-transfect K562 cells with 45% efficiency, while maintaining nearly 100% cell viability.

PACS or OCIS codes

## Introduction

Advancements in precision and control within biotechnology and nanomedicine, including the discovery of novel drugs, the synthesis of genetic materials, and the development of other biomedical cargoes, have paved the way for clinically relevant applications.[1–8] However, spatiotemporal selective and efficient intracellular delivery of these payloads into the cellular cytoplasm and nuclei of specific populations of cells remains a barrier to their therapeutic implementation.[9,10] For example, while efficient, viral-mediated gene transfers are cell-specific, limited to the delivery of nucleic acids, often time intensive, immunogenic, and not inherently spatially selective.[11,12] Meanwhile, lipid-based DNA-complexing agents can be tailored to broaden the range of cargo and target cell types, but they are only moderately efficient and require a series of reagents for cell line-specific protocols.[13–16] Electroporation is a method that does enable direct delivery of genes into cells and the nucleus; however, it is hampered by risks of cell damage and low post-treatment viability, particularly in sensitive cell lines.[12,17,18] Additionally, although these are commercial methods of electroporation, they do not permit spatiotemporally selective delivery of multiple cargoes which is useful for targeted combination therapies and co-delivery of biomolecules and drugs.[19] High-precision, spatiotemporally selective delivery to cells can be achieved through techniques like optoporation, which employs femtosecond lasers to target individual cells, or nano-needle injections; however, without automation, these methods have limited throughput, as each cell must be targeted individually.[20–24]

In this paper, we address the above challenges by developing a system that enables efficient spatiotemporally selective delivery of carrier-free cargo. We formulate a biocompatible polymer composed of carbon-black-embedded polydimethylsiloxane (CB-PDMS), irradiated by a 1064-nm 11-ns pulsed neodymium-doped yttrium aluminum garnet; Nd:Y3Al5O12 (Nd:YAG) laser, which leads to a transient photothermal disruption of the cell membrane. After culturing adherent cells on these substrates, we can create two distinct cell populations within a single device, each containing a different fluorescently labeled cargo molecule. To extend the approach to suspension cells (*e.g.*, blood and immune cells), we fabricate microcuvette chambers designed to bring suspension cells into direct contact with the CB-PDMS substrate. Upon laser irradiation, we demonstrate successful delivery of fluorescently labeled Dextran (FITC-Dextran, 4, 10, 20, and 70 kDa), as well as cy3-siRNA (14 kDa) into the cytoplasm of K562 cells. This approach suggests new opportunities for *in vitro* cell biology research and for therapeutic applications involving RNA interference.[25–27]

**Results & Discussion**

*Carbon Black Embedded PDMS.* We selected polydimethylsiloxane (PDMS), a widely used material in biomedical applications, as a base substrate material due to its excellent biocompatibility, chemical stability, easy processability, gas permeability, optical transparency, and tunable mechanical properties.[28] However, PDMS lacks the ability for photothermal conversion on its own, a benefit for many bioimaging applications, as it does not inherently interact with biocompatible near-infrared lasers in the mJ range. To address this limitation, we incorporate within the PDMS matrix carbon black nanoparticles (CBNP), which are commonly used in suspension for laser-mediated drug delivery.[10,29–31] This approach enhances photothermal conversion while minimizing potential cytotoxicity, as the CBNPs do not come into direct contact with cells.

We prepare CB-PDMS substrates by mixing CBNPs (Strem Chemicals, acetylene carbon black, 50% compressed) with 10:1 (base:catalyst) Sylgard 184 (Dow Corning) and fabricated into either flat substrates by casting into petri dishes (Figure 1) or microcuvette chambers using standard photolithography techniques (Figure 2a).[32] Surface characterization using x-ray photoelectron spectroscopy (XPS) and scanning electron microscopy (SEM) confirms that the CBNPs are embedded consistently within the bulk of the PDMS and are not present on the surface of the substrate (Figure S1). Furthermore, we demonstrate that CB-PDMS substrates are biocompatible, as HeLa CCL-2 cells (ATCC) cultured on the substrates show high viability (see Supplementary Information for additional details).

Unlike transparent PDMS, CB-PDMS enables optothermal intracellular delivery of suspended cargo when irradiated by a 1064-nm 11-ns laser pulses at a 50-Hz repetition rate. By adjusting the concentration of carbon black, we observe a linear increase in 4-kDa FITC-Dextran delivery as CBNP loading in PDMS increases (Figure S2). However, compared to the flat casting technique, the soft lithography microcuvette fabrication can be more challenging, as the viscosity of the PDMS increases with CBNP concentration, we select a concentration of 2 wt% CBNP as a sufficient balance between efficient delivery and manageable fabrication for the microcuvette chambers (Figure S2).

*Delivery to Adherent Cells*. To demonstrate spatiotemporal intracellular delivery, we culture HeLa cells on flat 5 wt% CB-PDMS substrates. First, we replace the cell medium with a solution of Invitrogen™ Dextran, Cascade Blue™ ("dextran blue", 3 kDa, 25 mg/mL) in phosphate-buffered saline (PBS), using a programmed motorized stage to irradiate a 4 x4 mm square at a speed of 7.5 mm/s with a 1064-nm 11-ns pulsed laser (50 mJ/cm², 8 laser pulses per cell) (Figure 1a). After five minutes, we replace the dextran blue solution with a 0.57-mg/mL solution of calcein green (0.62 kDa, 0.57 mg/mL) in PBS, and we irradiate the sample with the programmed motorized stage tracing out the letter "H" (Figure 1b).

Fluorescence imaging, shown in Figure 1b-c, confirms successful sequential delivery of dextran blue and calcein green. We observe that some cells fluoresce in both the blue and green channels, suggesting that five minutes is insufficient for full membrane healing between deliveries. The cell detachment occurring in the laser-irradiated areas, indicated by the red-orange channel in Figure 1d, is likely due to lack of cell-adhesion on the substrate. To avoid this detachment, one could plasma treat the surface and/or coat the substrate surface with a cell adhesion promotor such as fibronectin or polydopamine.[33,34]

*Delivery to Suspension Cells.* For delivery to K562 (ATCC) suspension cells, we fabricate microcuvette chambers using imprint lithography of CB-PDMS. We prepare a positive mold of the microcuvette chambers (overall dimensions: 38 mm x 11 mm, individual channel width: 500 μm) using negative photoresist, and over this mold, we cast a thin 1-mm layer of CB-PDMS followed by a thicker supporting layer of clear PDMS. After the sample is cured, we use a 1 mm biopsy punch to create inlet and outlet holes and bind the chamber to glass to create the final device (Figure 2a). We prepare solutions of fluorescein isothiocyanate (FITC)-Dextran polysaccharide (4, 10, 20, and 70 kDa) at a concentration of 25 mg/mL in Leibovitz's L-15 medium. Using a programmed raster-scanning motorized stage, we then selectively irradiate the microcuvette chambers, limiting cargo delivery to cells in the path of the laser (Figure 2b-c). After irradiation, we image the cells with a confocal microscope to visualize cargo uptake (Figure 2c inset) and eject the content of the microcuvette chamber into a vessel by injecting 0.5 mL of Leibovitz's L-15 Medium into the inlet, followed by a 0.5-mL air injection and a final 0.5-mL of medium to fully evacuate the chamber. We then use flow cytometry of the collected sample to quantify delivery efficiency (Figure 3).

To determine how the channel height and laser irradiation fluence affects delivery efficiency, we fabricated multiple microcuvette chambers, varying the channel height from 10 µm to 25 µm, while keeping the width of the channel fixed at 500 µm. The average diameter of K562 cells is about 12 µm in this setup. At a channel height of 12 µm, the cells become slightly compressed in the horizontal direction, and their vertically measured diameter increases to 12.7 µm; at a channel height of 10 µm, the cell diameter increases to 14.3 µm (Figure S4a). For channel heights larger than 15 µm, the FITC-Dextran (4 kDa) delivery does not vary with channel height. Reducing the channel height from 15 µm to 12 µm increases the delivery rate by about 30%, indicating that increased contact with the substrate significantly improves the efficiency of photothermal energy transfer and therefore intracellular delivery (Figures S4b). At a channel height of 10 µm, the delivery efficiency increases to 48% compared to the channel height of 15 µm. Despite the compression of the K562 cells, their post-irradiation viability remains near 100%, indicating that the cell compression does not significantly affect cell survival. For a fixed channel height of 12 µm, we find that a fluence of 50 mJ/cm$^2$ optimizes delivery and ability to eject cells from the microcuvette chambers. At fluences above 50 mJ/cm², we visually observed an increase in delivery using confocal microscopy; however, we were unable to reproducibly eject the cells from the chambers, preventing their recovery for analysis (Figure S3). While this analysis focuses on K562 cells, different chamber heights and laser fluences may be required to optimize delivery in cells of varying sizes.

At a fixed channel height of 10 µm and 12 µm, the amount of FITC-dextran delivery is inversely related to the molecular weight of the cargo (Figure 3a). The effective hydrated radius of FITC-dextran in solution is roughly 1.4 nm for 4 kDa, 2.9 nm for 10 kDa, 4.3 nm for 20 kDa, and 7.1 nm for 70 kDa, implying that the small diameter of the pores created via this method leads to lower uptake of molecules with a larger effective hydrated radius.[35,36] At a laser irradiation fluence of 50 mJ/cm$^2$, viability is 100% — identical to a control group of cells that are neither injected into the chamber nor irradiated.

*Delivery of siRNA.* To illustrate the broader utility of this technology, we demonstrated the spatially selective intracellular delivery of delivered fluorescently labeled non-coding siRNA (ThermoFisher Silencer™ Cy™3-labeled Negative Control No. 1 siRNA) into K562 suspension cells. We diluted a stock solution of Cy3-siRNA (100 µM in nuclease-free water) at various concentrations (0.1, 1, 5, or 10 µM in Leibovitz's L-15 Medium), mixed this solution with K562 cells, and injected 3 µL of the resulting solution containing about 240,000 cells into a microcuvette chamber with a channel height of 12 µm. After irradiation, flow cytometry shows that increasing siRNA concentration leads to increased delivery: 12.2% at 0.1 µM, 29.3% at 1 µM, and approaching an upper limit of 45.9% at 5 µM and above (Figure 3b). Confocal microscopy qualitatively confirms siRNA delivery as a function of concentration (Figure 3b). It is important to note that even though the maximum tested concentration of siRNA (10 µM) is significantly lower than the concentration used for the 10-kDa FITC-Dextran (25 mg/ml which corresponds to 2.5 mM), the delivery of siRNA is 13% higher than that of 10-kDa FITC-Dextran. This is likely due to the smaller radius of siRNA compared to that of the 10-kDa FITC-Dextran (2 nm *versus* 2.9 nm, respectively), which permits easier diffusion through the transient pores formed in the cell membrane. As in the FITC-Dextran experiments, cell viability after siRNA delivery remains nearly 100% compared to an untreated control group.

**Conclusion**

We demonstrated that laser irradiation of CB-PDMS substrates enables spatiotemporally selective intracellular delivery of multiple cargo types to adherent and to suspension cells. Embedding carbon black particles into the bulk of a PDMS creates a biocompatible material that prevents particle contamination in cell cultures, allowing for distinct cargo delivery to different cell populations in a single device via targeted laser irradiation. CB-PDMS can be easily integrated into standard cell culture dishes and well plates through casting.

The use of CB-PDMS in microcuvette chambers demonstrates the versatility and ease of fabrication of this composite material. Laser-irradiated K562 cells in these chambers show no reduction in viability compared to untreated cells, confirming that the approach demonstrated in this paper is scalable and suitable for transfecting limited quantities of patient-isolated cell lines.

Spatially and temporally controlled intracellular delivery systems hold significant promise across a range of applications. In heterogeneous mixtures of cells, such as pancreatic cancer cells, precise delivery methods can target specific cell subpopulations and can enable effective studies of cellular interactions and responses within complex tissues. This level of control is especially useful for therapies requiring selective treatment, such as targeted gene editing or cellular reprogramming. In regenerative medicine, the ability to deliver large cargo, such as plasmids, CRISPR components, or signaling molecules directly into stem cells can greatly enhance their potential for selective cell differentiation and therefore complex tissue regeneration by enabling precise control over differentiation pathways from an otherwise homogenous population of isolated stem cells.

In the future, implanted or inserted CB-PDMS devices could be utilized at point-of-care in live patients in conjunction with targeted laser therapy. For example, therapeutic molecules, such as chemotherapeutic agents, could be delivered to cancerous or diseased cells within a tissue. Moreover, for imaging applications like magnetic particle imaging, transfecting cells with particles using spatially selective methods could allow for high-resolution tracking of cellular movements and distribution within the body. Additionally, achieving efficient delivery of cargo into hard-to-transfect immune cells opens new avenues for immunotherapy and gene-based treatments. These advancements in spatial and temporal delivery have the potential to transform the precision and efficacy of therapeutic interventions, diagnostics, and tissue engineering.

**Acknowledgments**

W.S. conceived the basic idea for this work. W.S. and A.C. designed and carried out the experiments, analyzed, and interpreted the results. N.B. designed the carbon black embedding protocol and interpreted the results. E.M. and D.W. supervised the research and the development of the manuscript. W.S. wrote the first draft of the manuscript. G.P carried out experiments related to materials characterization and interpreted results; all authors subsequently took part in the revision process and approved the final copy of the manuscript. The authors thank Zachary Niziolek for help with the flow cytometry measurements and Dr. Alex Raymond and Dr. Anna Shneidman for discussions.

The research described in this paper was supported by National Science Foundation under contract PHY-1219334. This work was performed in part at the Harvard University Center for Nanoscale Systems (CNS); a member of the National Nanotechnology Coordinated Infrastructure Network (NNCI), which is supported by the National Science Foundation under NSF award numbers ECCS-1541959 and ECCS-2025158. Part of the work was also performed at Harvard's Materials Research Science and Engineering Center (DMR-1420570). W.S. was funded by the American Dissertation Fellowship from American Association of University Women.

**Figure Captions**

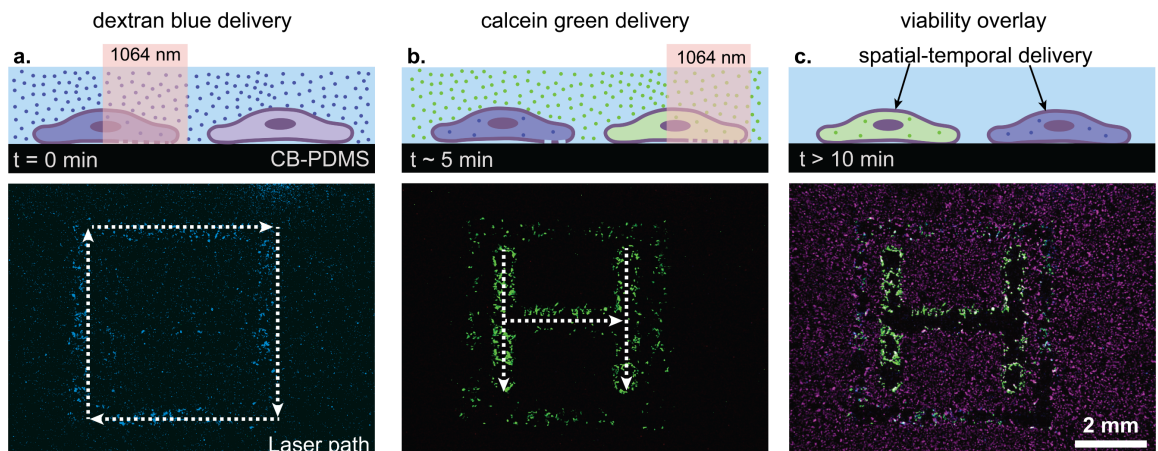

**Figure 1**: Spatiotemporal selective delivery of two fluorescently labelled cargoes to adherent HeLa cells. (a) Method for delivering fluorescently labelled cargoes to adherent HeLa cells on 5 wt% carbon-black-embedded PDMS (CB-PDMS) using 1064-nm ns laser pulses. (b) After the first laser irradiation the green cells show delivery of calcein green (0.62 kDa). (c) After 5 minutes, the laser is scanned again, and blue fluorescing cells show delivery of cascade blue Dextran (3 kDa). An overlay of the green, blue, and red (calcein AM red-orange viability stain fluorescence altered to be magenta instead of red).

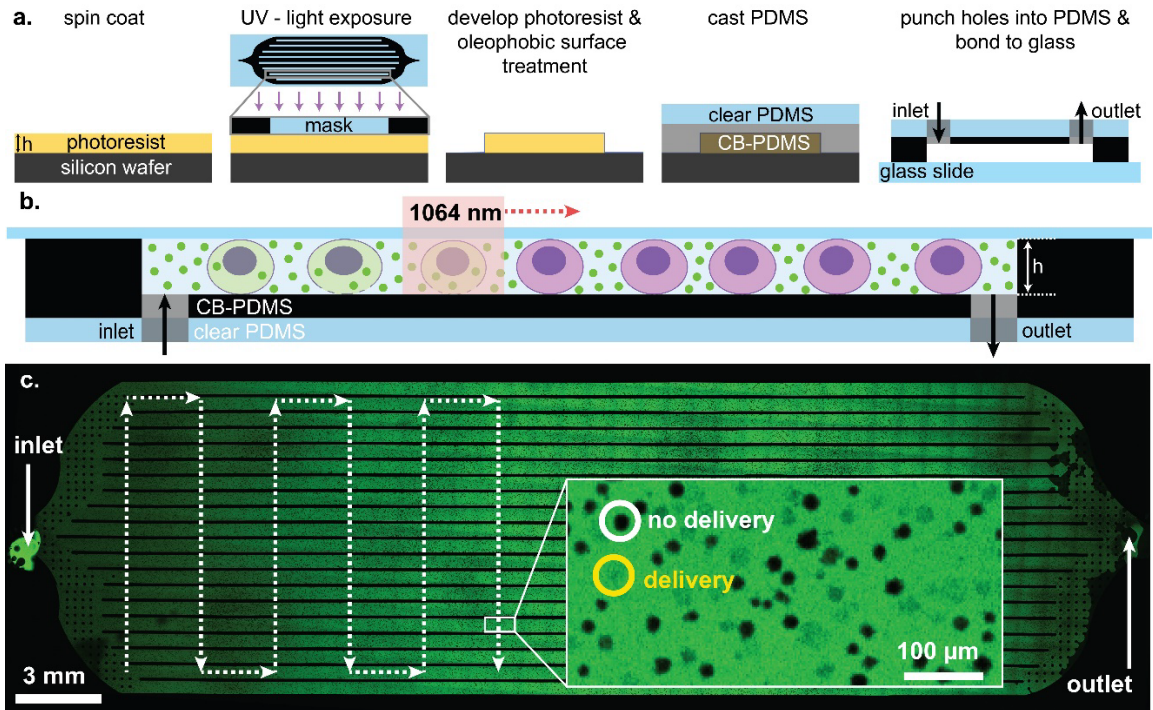

**Figure 2**: (a) Schematic of the soft lithography fabrication of microcuvette chambers. (b) Cross-section of device showing cells in a channel; $h$ = height of the channel. (c) Tile scan of confocal fluorescent images showing top view of a single microcuvette chamber ($h$ = 12 μm) containing FITC-Dextran solution (4 kDa, 25 mg/mL) and K562 suspension cells. Inset: After laser irradiation, cells that have taken up fluorescent cargo appear green. Walls and pillars supporting the microcuvette device appear black, as do cells to which the fluorescent cargo has not been delivered. The 2-mm wide dark areas are denser regions of cells.

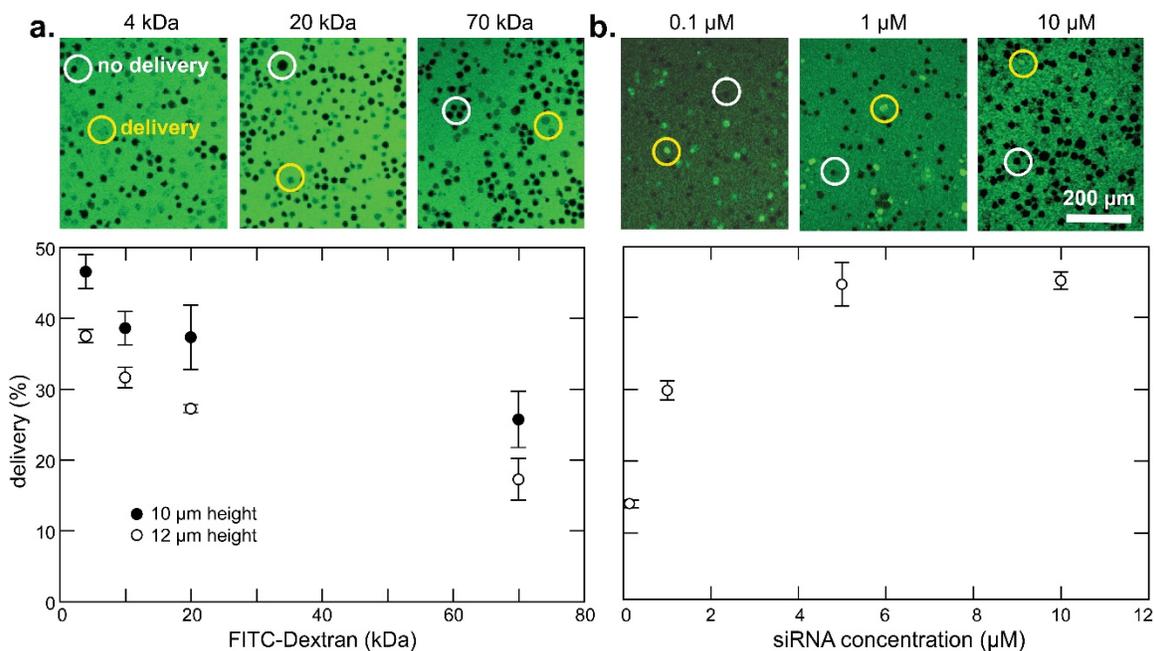

**Figure 3**: Delivery of (a) FITC-Dextran and (b) Cy3-siRNA to hard-to-transfect K562 cells. The confocal images (top) show sample areas that were irradiated with 1064-nm 11-ns laser pulses, showing cells that did and that did not take up cargo. Bottom: Flow cytometry measurements of cargo delivery. As the molecular weight of FITC-Dextran increases, delivery decreases for both 10- and 12-μm channel heights. As the concentration of Cy3-siRNA increases, delivery increases, leveling off around 45.9%.

# References

(1) Nicolas, J.; Mura, S.; Brambilla, D.; Mackiewicz, N.; Couvreur, P. Design, Functionalization Strategies and Biomedical Applications of Targeted Biodegradable/Biocompatible Polymer-Based Nanocarriers for Drug Delivery. *Chem Soc Rev* **2013**, *42* (3), 1147–1235. https://doi.org/10.1039/C2CS35265F.
(2) Wang, L. L.-W.; Gao, Y.; Feng, Z.; Mooney, D. J.; Mitragotri, S. Designing Drug Delivery Systems for Cell Therapy. *Nat. Rev. Bioeng.* **2024**, *2* (11), 944–959. https://doi.org/10.1038/s44222-024-00214-0.
(3) Magnani, M. Drug Delivery and Targeting System. *Emerg. Ther. Targets* **1998**, *2* (1), 145–146. https://doi.org/10.1517/14728222.2.1.145.
(4) Kim, B. Y. S.; Rutka, J. T.; Chan, W. C. W. Nanomedicine. *N. Engl. J. Med.* **2010**, *363* (25), 2434–2443. https://doi.org/10.1056/NEJMra0912273.
(5) Langer, R. New Methods of Drug Delivery. *Science* **1990**, *249* (4976), 1527–1533. https://doi.org/10.1126/science.2218494.
(6) Sanhai, W. R.; Sakamoto, J. H.; Canady, R.; Ferrari, M. Seven Challenges for Nanomedicine. *Nat. Nanotechnol.* **2008**, *3* (5), 242–244. https://doi.org/10.1038/nnano.2008.114.
(7) Wagner, V.; Dullaart, A.; Bock, A.-K.; Zweck, A. The Emerging Nanomedicine Landscape. *Nat. Biotechnol.* **2006**, *24* (10), 1211–1217. https://doi.org/10.1038/nbt1006-1211.
(8) Sansing, L. H.; Harris, T. H.; Welsh, F. A.; Kasner, S. E.; Hunter, C. A.; Kariko, K. Toll-like Receptor 4 Contributes to Poor Outcome after Intracerebral Hemorrhage. *Ann. Neurol.* **2011**, *70* (4), 646–656. https://doi.org/10.1002/ana.22528.
(9) Morshedi Rad, D.; Alsadat Rad, M.; Razavi Bazaz, S.; Kashaninejad, N.; Jin, D.; Ebrahimi Warkiani, M. A Comprehensive Review on Intracellular Delivery. *Adv. Mater.* **2021**, *33* (13), 2005363. https://doi.org/10.1002/adma.202005363.
(10) Stewart, M. P.; Langer, R.; Jensen, K. F. Intracellular Delivery by Membrane Disruption: Mechanisms, Strategies, and Concepts. *Chem. Rev.* **2018**, *118* (16), 7409–7531. https://doi.org/10.1021/acs.chemrev.7b00678.
(11) Bulcha, J. T.; Wang, Y.; Ma, H.; Tai, P. W. L.; Gao, G. Viral Vector Platforms within the Gene Therapy Landscape. *Signal Transduct. Target. Ther.* **2021**, *6* (1), 53. https://doi.org/10.1038/s41392-021-00487-6.
(12) Al-Dosari, M. S.; Gao, X. Nonviral Gene Delivery: Principle, Limitations, and Recent Progress. *AAPS J.* **2009**, *11* (4), 671. https://doi.org/10.1208/s12248-009-9143-y.
(13) Ganta, S.; Devalapally, H.; Shahiwala, A.; Amiji, M. A Review of Stimuli-Responsive Nanocarriers for Drug and Gene Delivery. *J. Controlled Release* **2008**, *126* (3), 187–204. https://doi.org/10.1016/j.jconrel.2007.12.017.
(14) Mora-Huertas, C. E.; Fessi, H.; Elaissari, A. Polymer-Based Nanocapsules for Drug Delivery. *Int. J. Pharm.* **2010**, *385* (1–2), 113–142. https://doi.org/10.1016/j.ijpharm.2009.10.018.
(15) Patil, S. D.; Rhodes, D. G.; Burgess, D. J. DNA-Based Therapeutics and DNA Delivery Systems: A Comprehensive Review. *AAPS J.* **2005**, *7* (1), E61–E77. https://doi.org/10.1208/aapsj070109.

(16) Phillips, A. J. The Challenge of Gene Therapy and DNA Delivery. *J. Pharm. Pharmacol.* **2001**, *53* (9), 1169–1174. https://doi.org/10.1211/0022357011776603.

(17) Marti, G.; Ferguson, M.; Wang, J.; Byrnes, C.; Dieb, R.; Qaiser, R.; Bonde, P.; Duncan, M.; Harmon, J. Electroporative Transfection with KGF-1 DNA Improves Wound Healing in a Diabetic Mouse Model. *Gene Ther.* **2004**, *11* (24), 1780–1785. https://doi.org/10.1038/sj.gt.3302383.

(18) Neumann, E.; Schaefer-Ridder, M.; Wang, Y.; Hofschneider, P. H. Gene Transfer into Mouse Lyoma Cells by Electroporation in High Electric Fields. *EMBO J.* **1982**, *1* (7), 841–845. https://doi.org/10.1002/j.1460-2075.1982.tb01257.x.

(19) Sun, W.; Sanderson, P. E.; Zheng, W. Drug Combination Therapy Increases Successful Drug Repositioning. *Drug Discov. Today* **2016**, *21* (7), 1189–1195. https://doi.org/10.1016/j.drudis.2016.05.015.

(20) Elnathan, R.; Barbato, M. G.; Guo, X.; Mariano, A.; Wang, Z.; Santoro, F.; Shi, P.; Voelcker, N. H.; Xie, X.; Young, J. L.; Zhao, Y.; Zhao, W.; Chiappini, C. Biointerface Design for Vertical Nanoprobes. *Nat. Rev. Mater.* **2022**, *7* (12), 953–973. https://doi.org/10.1038/s41578-022-00464-7.

(21) He, G.; Hu, N.; Xu, A. M.; Li, X.; Zhao, Y.; Xie, X. Nanoneedle Platforms: The Many Ways to Pierce the Cell Membrane. *Adv. Funct. Mater.* **2020**, *30* (21), 1909890. https://doi.org/10.1002/adfm.201909890.

(22) Wu, T.-H.; Teslaa, T.; Kalim, S.; French, C. T.; Moghadam, S.; Wall, R.; Miller, J. F.; Witte, O. N.; Teitell, M. A.; Chiou, P.-Y. Photothermal Nanoblade for Large Cargo Delivery into Mammalian Cells. *Anal. Chem.* **2011**, *83* (4), 1321–1327. https://doi.org/10.1021/ac102532w.

(23) Wu, Y.-C.; Wu, T.-H.; Clemens, D. L.; Lee, B.-Y.; Wen, X.; Horwitz, M. A.; Teitell, M. A.; Chiou, P.-Y. Massively Parallel Delivery of Large Cargo into Mammalian Cells with Light Pulses. *Nat. Methods* **2015**, *12* (5), 439–444. https://doi.org/10.1038/nmeth.3357.

(24) Tirlapur, U. K.; König, K. Targeted Transfection by Femtosecond Laser. *Nature* **2002**, *418* (6895), 290–291. https://doi.org/10.1038/418290a.

(25) Dana, H.; Chalbatani, G. M.; Mahmoodzadeh, H.; Karimloo, R.; Rezaiean, O.; Moradzadeh, A.; Mehmandoost, N.; Moazzen, F.; Mazraeh, A.; Marmari, V.; Ebrahimi, M.; Rashno, M. M.; Abadi, S. J.; Gharagouzlo, E. Molecular Mechanisms and Biological Functions of siRNA. *Int. J. Biomed. Sci. IJBS* **2017**, *13* (2), 48–57.

(26) Hu, B.; Zhong, L.; Weng, Y.; Peng, L.; Huang, Y.; Zhao, Y.; Liang, X.-J. Therapeutic siRNA: State of the Art. *Signal Transduct. Target. Ther.* **2020**, *5* (1), 101. https://doi.org/10.1038/s41392-020-0207-x.

(27) Devi, G. R. siRNA-Based Approaches in Cancer Therapy. *Cancer Gene Ther.* **2006**, *13* (9), 819–829. https://doi.org/10.1038/sj.cgt.7700931.

(28) Wolf, M. P.; Salieb-Beugelaar, G. B.; Hunziker, P. PDMS with Designer Functionalities—Properties, Modifications Strategies, and Applications. *Prog. Polym. Sci.* **2018**, *83*, 97–134. https://doi.org/10.1016/j.progpolymsci.2018.06.001.

(29) Pulskamp, K.; Diabate, S.; Krug, H. Carbon Nanotubes Show No Sign of Acute Toxicity but Induce Intracellular Reactive Oxygen Species in Dependence on Contaminants. *Toxicol. Lett.* **2007**, *168* (1), 58–74. https://doi.org/10.1016/j.toxlet.2006.11.001.

(30) López-Lugo, J. D.; Pimentel-Domínguez, R.; Benítez-Martínez, J. A.; Hernández-Cordero, J.; Vélez-Cordero, J. R.; Sánchez-Arévalo, F. M. Photomechanical Polymer Nanocomposites for Drug Delivery Devices. *Molecules* **2021**, *26* (17), 5376. https://doi.org/10.3390/molecules26175376.

(31) Kumar, S.; Li, A.; Thadhani, N. N.; Prausnitz, M. R. Optimization of Intracellular Macromolecule Delivery by Nanoparticle-Mediated Photoporation. *Nanomedicine Nanotechnol. Biol. Med.* **2021**, *37*, 102431. https://doi.org/10.1016/j.nano.2021.102431.

(32) Qin, D.; Xia, Y.; Whitesides, G. M. Soft Lithography for Micro- and Nanoscale Patterning. *Nat. Protoc.* **2010**, *5* (3), 491–502. https://doi.org/10.1038/nprot.2009.234.

(33) Lee, J. N.; Jiang, X.; Ryan, D.; Whitesides, G. M. Compatibility of Mammalian Cells on Surfaces of Poly(Dimethylsiloxane). *Langmuir* **2004**, *20* (26), 11684–11691. https://doi.org/10.1021/la048562+.

(34) Toworfe, G. K.; Composto, R. J.; Adams, C. S.; Shapiro, I. M.; Ducheyne, P. Fibronectin Adsorption on Surface-activated Poly(Dimethylsiloxane) and Its Effect on Cellular Function. *J. Biomed. Mater. Res. A* **2004**, *71A* (3), 449–461. https://doi.org/10.1002/jbm.a.30164.

(35) Kumar, S.; Li, A.; Thadhani, N. N.; Prausnitz, M. R. Optimization of Intracellular Macromolecule Delivery by Nanoparticle-Mediated Photoporation. *Nanomedicine Nanotechnol. Biol. Med.* **2021**, *37*, 102431. https://doi.org/10.1016/j.nano.2021.102431.

(36) Arrio-Dupont, M.; Cribier, S.; Foucault, G.; Devaux, P. F.; d'Albis, A. Diffusion of Fluorescently Labeled Macromolecules in Cultured Muscle Cells. *Biophys. J.* **1996**, *70* (5), 2327–2332. https://doi.org/10.1016/S0006-3495(96)79798-9.

**Carbon black embedded PDMS**

To confirm that the carbon black particles are embedded within Sylgard 184 and not presenting at the surface, characterization is done using x-ray photoelectron spectroscopy (XPS, Thermo Scientific K-Alpha Surface Analyzer) and scanning electron microscopy (Zeiss Gemini 360 FE-SEM).

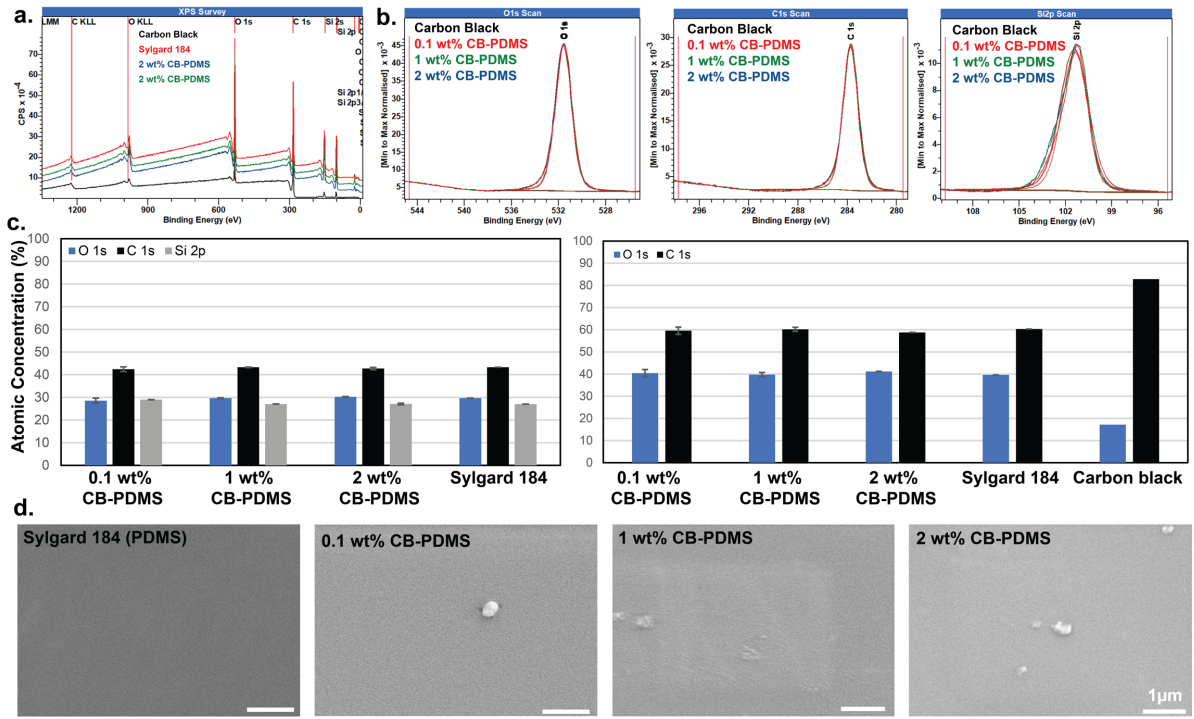

Figure S1: Surface characterization of CB-PDMS. (a) XPS survey scans comparing neat carbon black, neat PDMS and CB-PDMS. (b-c) High resolution scans of O 1s, C 1s and Si 2p peaks are measured to quantify the atomic concentrations at the surface. (d) SEM images show there are no discernible differences between PDMS and CB-PDMS.

The atomic concentration of O 1s, C 1s and Si 2p peaks were quantified from high resolution scans (Figure S1a-c). Compared to Sylgard 184 there is no significant difference between the atomic concentration on the surface. Additionally, neat carbon black shows a significantly different atomic concentration ratios than CB-PDMS, indicating that the particles are embedding. Furthermore, visually there are no discernible differences observed between the surfaces under SEM (Figure S1d).

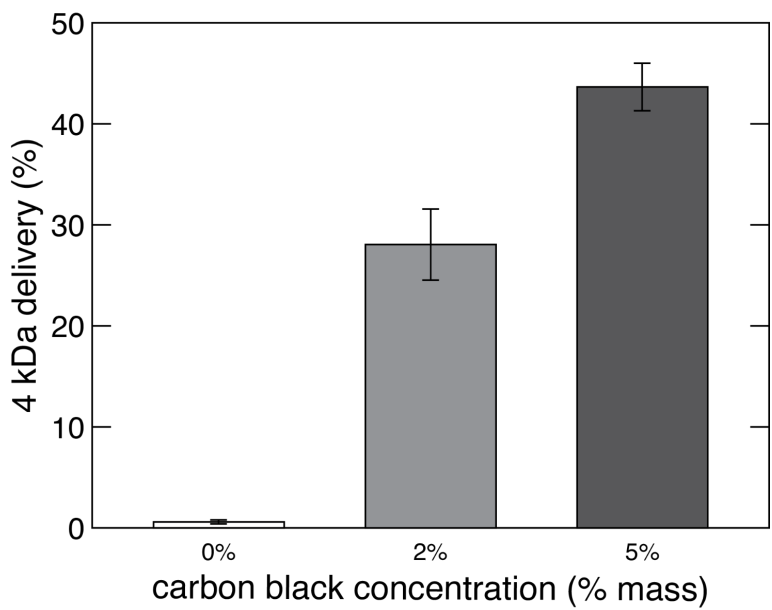

*Figure S2: Delivery percentage of 4 kDa FITC-Dextran to K562 cells for 0, 2, and 5% CB-PDMS microcuvette chambers (h = 12 μm)*

To determine the optimal concentration of carbon black we tested the delivery of 4 kDa FITC-Dextran to K562 cells in microcuvette chambers. Although flow cytometry shows that delivery rate increases with increased carbon black concentration, the viscosity at 5% carbon black makes the molding against small features during the soft lithography process challenging.

**Fabrication of flat carbon black embedded PDMS**

To prepare flat CB-PDMS substrates, 5wt% carbon black nanoparticles (VWR) are mixed with 1:10 Sylgard 184 (Dow Corning) and the mixture is poured into a 100-mm Petri dish. The mixture is degassed for 15 minutes in a vacuum chamber to remove air bubbles, and then transferred into an oven at 65°C and cured for a minimum of four hours. Finally, to promote cell growth upon cell culture, the surface of the CB-PDMS is oxygen plasma treated to create a hydrophilic surface.

**Fabrication of carbon black embedded PDMS microcuvette chambers**

Soft Lithography is used to fabricate the microcuvette chambers for intracellular delivery to suspension cells (Figure 2a). The process is parts is as follows: Negative photoresist (SU-8) is spin coated onto a silicon wafer (University Wafers). The viscosity and spin speed determines the thickness of the photoresist and, in turn, the height of the final microcuvette chamber. The coated silicon wafer is then baked at 65°C and 95°C on a hot plate to stabilize the photoresist. Next, a photomask designed in AutoCAD with the pattern of the microcuvette chambers is placed over the photoresist-coated wafer, and the wafer is irradiated with UV light. The coated wafer is then baked again at 65°C and 95°C on a hot plate to ensure complete curing. Propylene glycol methyl ether acetate (PGMEA) is used as a photoresist developer to dissolve the unexposed areas to UV light, leaving behind the mold's inverse structures of the microcuvettes. Finally, to prevent adhesion between the PDMS and the mold, the surfaces of the cured photoresist and silicon wafer are modified with 0.1% Trichloro(1H,1H,2H,2H-perfluorooctyl)silane in Novec™ HFE 7500 (3M) to create an oleophobic surface. The CB-PDMS (2wt% carbon black) mixture is poured over the silicon master, forming a 1- to 2-mm layer that ensures all microfeatures on the wafer are fully covered. The wafer is then placed in a vacuum chamber for at least 15 minutes to remove any air bubbles, followed by curing at 65°C until set. After curing, a second, thicker layer of clear PDMS is poured on top to form a composite slab approximately 5 mm thick. The sample is degassed under vacuum for 5 to 10 minutes, then cured in an oven at 65°C for at least four hours. Once cured, the two-layer PDMS slab is carefully peeled off the silicon master using a scalpel. The PDMS slab contains the inverse structures are from those of the silicon master. Next, a 1-mm biopsy punch is used to create inlet and outlet holes for each device in the PDMS slab. Finally, the PDMS slab is bonded to a 2-mm thick glass slide so the entire microcuvette chamber is enclosed except for its inlet and outlet. To improve the adhesion between the glass slide and CB-PDMS, both are oxygen plasma treated before being pressed together. To make the interior of the chambers hydrophobic, Aquapel is injected and then immediately flushed out using air; any residual Aquapel is evaporated by baking the sample for half an hour at 65 °C. [39]

**HeLa Cell Culture Protocol**

For experiments on flat substrates, HeLa cells (ATCC) are cultured in standard cell medium (Dulbecco's Modified Eagle Medium supplemented with 10% Fetal Bovine Serum and 1% Penicillin-Streptomycin). The cells are cultured at 37 °C in the presence of 5% carbon dioxide and maintained at passaged at ~80% confluency.

**K562 Cell Culture Protocol**

For the experiments in the microcuvette chamber we use K562 cells (ATCC) are cultured in 500 mL RPMI-1640, 50 mL FBS, 5 mL 200 mM L-Glutamine, and 5 mL Penicillin-Streptomycin. The cells are cultured at 37 °C in the presence of 5% carbon dioxide and maintained at a cell culture density of 0.2 to 1 x $10^6$ cells/mL.

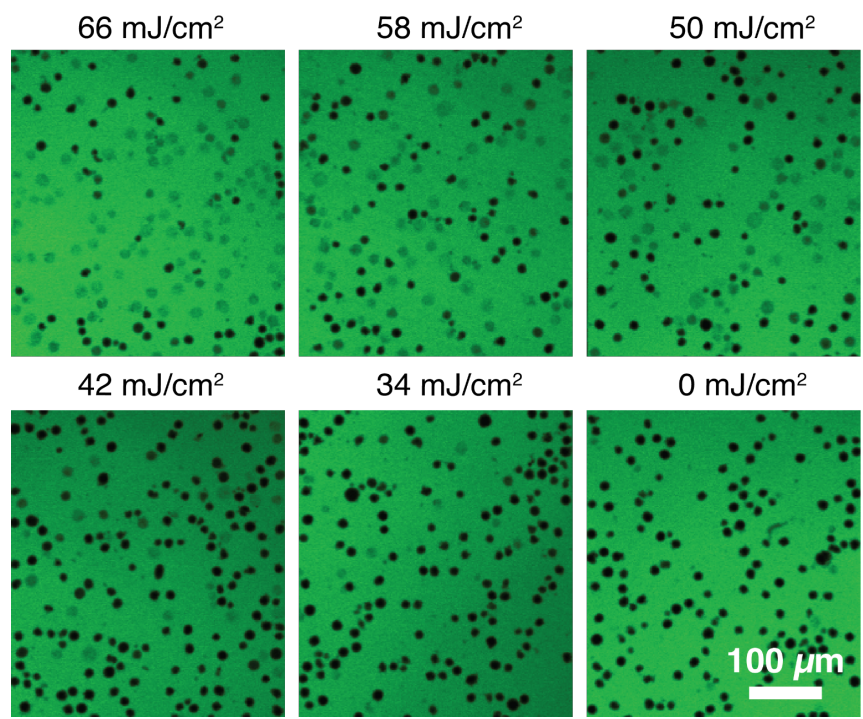

***Figure S3:*** *Pose laser irradiation fluorescence images for varying laser fluences in the presence of 4 kDa FITC-Dextran in 12 μm microcuvette chambers.*

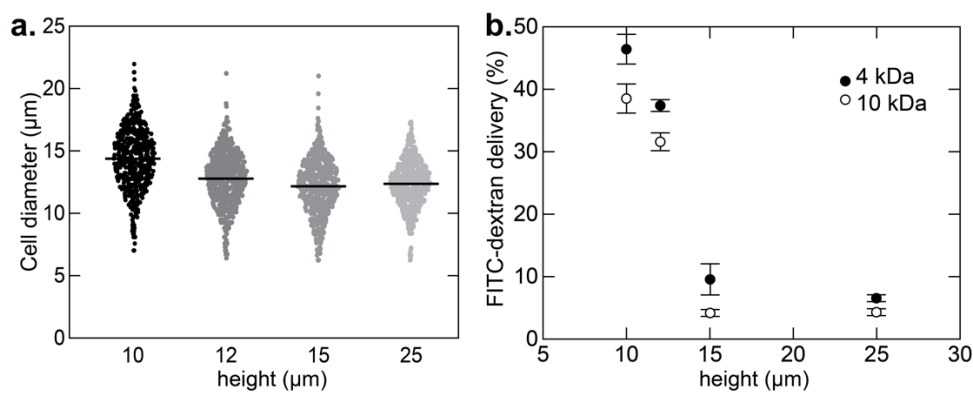

***Figure S4:*** *(a) Cell diameter in varying chamber heights. (b) FITC-dextran delivery using* 50 mJ/cm².

**Laser Parameters**

    A 1064-nm nanosecond pulsed laser (Quantel, Ultra20). The number of laser pulses applied to each cell are controlled by adjusting the stage scan speed where $\# \, pulses = \frac{\emptyset \times f_{rep}}{v}$. The $\emptyset$, beam

diameter is measured using a coherent beam profiler, the repetition rate of the laser is fixed at 50 Hz and the stage scan speed is adjusted. The fluence (mJ/cm²) applied to the substrate is adjusted but attenuating the beam power. The fluence can be calculated as $Fluence = \frac{P_{avg}}{\emptyset}$. The laser optimal laser fluence is 50 mJ/cm². This value optimizes the delivery of FITC-Dextran and the ability to eject the cells from the chambers. Above this fluence, the cells tend to remain trapped in the chambers making any quantification of delivery challenging (Figure S3).

**Delivery of fluorescent labelled cargo to adherent HeLa cells on flat substrates**

Flat 5wt% CB-PDMS oxygen plasma treated substrates are placed at the bottom of 100-mm petri dishes and are sanitized with 70% ethanol. To prevent detachment, double sided Kapton tape is used to keep the substrate in place. HeLa cells are seeded at a density of 1x10⁶ cells/mL in cell media and enough cell suspension solution is added to cover the substrate. The cells are allowed to adhere and grow for 24h. Following this, the substrate with the cells is gently removed using tweezers and transferred into a new petri dish. Immediately, the calcein green (0.62 kDa) solution in Leibovitz medium (0.57 mg/mL) is added to the petri dish. The sample is placed in the beam of a 11-ns pulsed laser with a 1064-nm wavelength laser and a repetition rate of 50 Hz. The Gaussian beam is 1.2-mm and held constant and the laser excitation fluence is 50 mJ/cm². The stage on which the sample is placed translates the sample across the laser beam at a speed of 7.5 mm/s. After the first scan, the calcein green solution is removed from the Petri dish and replaced with a solution of Cascade Blue-Dextran (3 kDa, ThermoFisher Scientific). The substrate is laser-scanned again in a different pattern. Afterwards, the cells on the substrate are stained with viability indicator, calcein AM red-orange, washed with PBS, and the entire substrate is imaged in a widefield upright fluorescence microscope to visualize the delivery of calcein green and Dextran blue.

**FITC-dextran and Cy3-siRNA delivery to K562 cells using microcuvette chambers**

Solutions of Fluorescein isothiocyanate (FITC)-Dextran polysaccharide molecules (excitation max = 499 nm, emission max = 517 nm) with molecular weights ranging from 4 to 70 kDa are prepared at a concentration of 25 mg/mL in Leibovitz's L-15 medium. K562 cells are centrifuged at 1200 rpm for 5 minutes at room temperature. The supernatant medium is removed with a pipette, and the pellet of cells at the bottom of the centrifuge tube is then resuspended with the cargo-containing Leibovitz's L-15 Medium to obtain a final cell concentration of 80 million cells/mL. To insert the cells into the microcuvette chambers, a suspended cell sample is loaded into a syringe with a blunt needle and injected through an inlet hole until the whole microfluidic chamber is visibly filled with liquid (about 3 µL containing 240,000 cells for a microfluidic chamber of 10 µm in height). To deliver the cargo to the cells, the glass slide with the microfluidic chambers, filled cells and cargo, are placed on a raster scanning stage, in the laser beam path.

After irradiation, the cells are imaged with a confocal microscope (Leica SP5) to measure cargo uptake. To eject the cells from the microcuvette chambers, 0.5 mL of Leibovitz's L-15 Medium is injected into the inlet using a syringe needle and the ejected cells and medium are collected from the microcuvette chamber outlet into a collection vessel using plastic tubing fitted with an Eppendorf tube. Next, 0.5 mL of air is injected to further evacuate the chamber before a final 0.5 mL of medium is injected. The collected cells are then prepared for flow cytometry. After irradiation, the cells are imaged with a confocal microscope to measure cargo uptake. To eject the cells from the microfluidic chambers, 0.5 mL of Leibovitz's L-15 Medium is injected into the inlet using a syringe needle and the ejected cells and medium are collected from the microfluidic chamber outlet into a collection vessel using plastic tubing fitted with an Eppendorf tube. Next, 0.5 mL of air is injected to further evacuate the chamber be9fore a final 0.5 mL of medium is injected. The collected cells are then prepared for flow cytometry (10,000 cells are measured for each chamber and standard error is denoted).

For the siRNA experiments, we use fluorescently labelled siRNA (ThermoFisher Silencer™ Cy™3-labeled Negative Control No. 1 siRNA with excitation max = 547 nm, emission max = 563 nm), making bulk stocks at a concentration of 100 µM with nuclease-free water to prevent any nucleic acid degradation in storage. Before each experiment, we prepare samples at concentrations of 0.1, 1, 5, and 10 µM using Leibovitz's L-15 Medium. The preceding procedure is the same as above following this.

To minimize background signal from the undelivered labelled fluorescent molecules remaining in the medium, the solution retrieved from microcuvette chamber is spun down at 1200 rpm for 5 minutes. The supernatant is removed and the cell pellet is resuspended with 200 µL of Leibovitz's L-15 Medium and 1 µL of the fluorescent viability stain propidium iodide (1 mg/mL DMSO stock) and incubated at room temperature for a minute. To check the viability either propidium iodide (FITC-Dextran delivery) or DAPI stain (siRNA delivery). Viability is not reported as we observe that it is nearly 100%. For each set of experiments of varying cargo type and concentration, a population of cells in medium with cargo present but not loaded into a microchamber serves as a complete control. Each experiment for a cargo type and concentration, laser fluence, and chamber height is repeated three times and to correct for any background cargo delivery that happens during cell transport in the chamber, each experiment for a given cargo type and concentration is repeated without laser illumination.